# Who Said Only Military Officers Can Deal with Uncertainty? On the Importance of Uncertainty in EdTech Data Visualisations

Felicitas Macgilchrist[a] and Juliane Jarke[b]

[a] Re:Lab & Department of Educational Sciences, Carl von Ossietzky University of Oldenburg, Germany, felicitas.macgilchrist@uni-oldenburg.de
[b] BANDAS Center & Department of Sociology, University of Graz, Austria

**Abstract**
AI-powered predictive systems have high margins of error. However, data visualisations of algorithmic systems in education and other social fields tend to visualise certainty, thus invisibilising the underlying approximations and uncertainties of the algorithmic systems and the social settings in which these systems operate. This paper draws on a critical speculative approach to first analyse data visualisations from predictive analytics platforms for education. It demonstrates that visualisations of uncertainty in education are rare. Second, the paper explores uncertainty visualisations in other fields (defence, climate change and healthcare). The paper concludes by reflecting on the role of data visualisations and un/certainty in shaping educational futures. It also identifies practical implications for the design of data visualisations in education.

**Keywords:**
Algorithms, Dashboards, Datafication, Data Subjects, Data Visualization, Platforms, Predictive Analytics, Uncertainty

**Introduction**
The algorithmic systems of today's datafied world have been diagnosed as providing a 'utopia of certainty' that prioritises 'the promise of guaranteed outcomes' (Zuboff, 2019, p. 398, 497). At the same time, these systems have high margins of error. They are 'brittle' and tend to 'break in ways that have more harmful impacts on people who are already vulnerable' (McQuillan et al., 2024, p. 2; see Baker & Hawn, 2022; Eynon, 2024; Knox, 2022). When the algorithmic systems ('AI' or not) across education, health, social work, military, policing and other social fields visualise data *as if they were* accurate, exact or certain, they render invisible the underlying approximations and uncertainties of both the algorithmic systems and the social settings in which these systems operate. Such visualisations stand in stark contrast to the lived experiences of people working in these domains (Whitman, 2020). They also stand in contrast to the very idea of education for many theorists who see education as inherently uncertain and open-ended (Britzman, 2009; Allert et al., 2017). Yet data visualisations that suggest certainty lie at the heart of many decisions made about students' lives. Data visualisations showing, for instance, students to be at high risk of not persisting at college, lead to an educator thinking they should advise these students away from post-secondary education (Hartong &

Förschler, 2019). Data visualising a race between technology and education aim to impact educational governance and the procurement of technology (Mikhaylova & Pettersson, 2024). A data visualisation of equitable classroom participation encourages a teacher to pose fewer questions to minoritised students (Reinholz & Shah, 2021). At stake in data visualisations of un/certainty are, thus, core ethicopolitical issues about how datafied education impacts students' lives.

Critical algorithm studies have raised questions about who produces and circulates (valid) knowledge in algorithmic regimes, how truth claims are made, and how high-stakes decisions are taken (Jarke et al., 2024). In this paper, we argue that data visualisations of certainty in educational technologies (edtech) play a crucial role in devaluing 'situated knowledges' (Haraway, 1988) that embrace the uncertainty and open-endedness of education and, more generally, the world.

The argument draws on a critical speculative research design (Ross, 2023; Wilkie et al., 2017). In the approach developed here, Phase 1 analyses a corpus of 19 existing data visualisations from learning management systems (LMS) which utilise predictive analytics. We investigate if/how they visualise uncertainty in their predictions of student success. Findings show that data visualisations of uncertainty in education are exceedingly rare. Phase 2 thus speculates on modes of visualising uncertainty that shift contemporary 'aesthetic practices' (Ratner & Ruppert, 2019) for presenting data in education. A key finding is that some other domains (defence, climate change and healthcare) expect professionals *to be able to–and in most instances to need to*–deal with uncertainty in order to perform their work well. The paper concludes by reflecting on the educational implications of data visualisations, and the need for a third phase of this research in which designers and educators would co-design data visualisations that embrace uncertainty.

**The importance of uncertainty in edtech data visualisations**
In this section, we review literature that demonstrates the increasingly central role of data visualisation for professional practice in education. We then relate this work to literature on the inherent uncertainty of education. Third, we review literature on engaging with uncertainty visualisations in and beyond education.

*Data Visualisation in Education*
Research on the increased datafication of education has pointed to the role of decontextualised, simplified and quantified data, especially digital data, in contemporary modes of organising and governing schools. Drawing on, inter alia, research on the social construction of statistics, computational models and algorithms, these studies show how today's forms of data render teaching a managerial practice, increase competition, responsibilise students, produce reductive accounts of what counts as 'good education' and increase inequalities among students (e.g. Bock et al., 2023; Bradbury, 2019; Decuypere & Landri, 2021; Perrotta & Williamson 2018; Selwyn et al., 2022; Thompson & Prinsloo, 2023). A key aspect is the visualisation of data in graphs, charts and other diagrams. The educational technology (edtech) industry markets its products as offering 'actionable data' which aid decision-making. These data visualisations are not, however, neutral representations of 'facts', but intensely political: they 'constitute a complex sociotechnical act involving a variety of actors and technologies with the persuasive power to shape people's engagement and interaction with the world itself' (Williamson, 2016, p. 132). Their persuasive power stems not only from the rational power of numbers or the logic of quantification, but also from the affective impact of data visualisations (Kennedy & Hill, 2018). Entangling the numeric and the visual, data visualisations evoke a range of emotions among users. They tell stories, weaving the numerical with the argumentative and the affective (Jarke & Macgilchrist, 2021). They invite learners to relate to themselves in specific self-monitoring ways (Vanermen et al., 2024) and impact how teachers see students (Ratner, 2024). They shape anticipations about the future and make worlds by rendering some futures more visible than others (Madsen, 2024).

Part of the persuasive power of data visualisations stems, we suggest, from the clarity and apparent unambiguity of their output. The dashboard of Brightspace, for instance, a leading LMS described by its CEO as including 'the next generation of artificial intelligence and machine learning', includes visualisations of data representing engagement, success and social learning (Jarke & Macgilchrist, 2021). Each of these visualisations operates with clear colour codes, bars, data points or charts, suggesting that the LMS offers clear insights into students' learning, helpful for predicting their future. Moodle, another dominant LMS worldwide, uses up to 49 indicators, also called 'predictors' in the Moodle documentation, in its algorithmic visualisation process.[1] The dashboard data visualisations are based on a host of proxies for cognitive and social engagement, including when students accessed the course, if they have written and saved content on the site, and how often they interact on the site with other participants. Indicators for these activities are created and connected before being processed into simple data visualisations.

Ethnographic research has identified the 'aesthetic practices' as statisticians and others create these kinds of data visualisations. When 'data cleaning', for instance, they identify, correct or delete 'absence, inaccuracy, and indeterminacy' in datasets (Ratner & Rupert, 2019, p. 11f). In the final visualisations – e.g., heat maps or bar charts – these indeterminacies and other frictions are no longer visible.

***Uncertainty in Education***
However, precisely such indeterminacies, frictions and uncertainties have been described as constitutive for educators' professional practice (Cramer et al., 2019). Educational theory generally agrees on the need to accept the inevitable contingency of classroom practices in order to make decisions, act and react in dynamic situations in which we need to expect the unexpected (Combe et al., 2018). This is the 'beautiful risk' of an education beyond rote learning, i.e., when education is seen as inherently contingent and open-ended (Biesta, 2013). This ontological dimension makes teaching an 'impossible profession' (Britzman, 2009). If the design of pedagogical interventions is not understood as an engineering task, in which problems are first identified and then solutions developed, but as an iterative and 'materially mediated process where problem and solution are co-evolutionarily entangled' (Macgilchrist et al., 2024), then we are dealing with ontological uncertainty.

Beyond ontological uncertainty, Bonnet and Glazier (2023) identify four types of uncertainty in education: (i) disciplinary uncertainty, the always already preliminary nature of scholarly knowledge, which changes over time, (ii) curricular uncertainty, the extent to which teachers adopt a role as 'brokers of uncertainty' (ibid., p.4) and discuss the limits of disciplinary knowledge with their students, (iii) pedagogical uncertainty, the unpredictability of classroom interactions and the impossibility of truly knowing if students have learned, and (iv) contextual uncertainty, the shifting policies, economic mandates and technological changes, to which educational systems (and teachers) feel they must react (Gilead & Dishon, 2022). Although there are few empirical studies of uncertainty in education, in a study asking teachers about classroom practices, *all* interviewees could relate experiences of pedagogical uncertainty (Hinzke et al., 2021). In some situations, they felt they did not live up to their expectations, and the uncertainty was experienced as stressful. In other situations, they felt unprepared, but enjoyed the dynamics of reacting flexibly to students' spontaneous interactions. Reports from practitioners thus align with the ontological commitments to the uncertainty of being that are prevalent in critical or (post-)humanist theories of education. This perspective is the backdrop for this article's inquiry into how uncertainty is made relevant or not in educational dashboards and systems, with implications for how educators may engage with un/certainty in their professional practice.

### *Engaging with Uncertainty Visualisations*

When an educational data dashboard in Denmark visualised uncertainty, teachers in a pilot study initially misunderstood the visualisation (Ratner & Bundsgaard, 2017). Rather than further developing, refining and clarifying the uncertainty visualisation, the design team quickly dropped it. Indeed, removing uncertainty from visualisations can be beneficial for a number of reasons, e.g., because uncertainty is complex, because uncertainty information is presented in different ways, because uncertainty requires an additional visual dimension (from 2D to 3D, or with animations, overlay, sounds or colours), because uncertainty can appear to be the most important aspect of the visualisation, because of a desire for consistency, or because calculations with uncertain data further propagate the uncertainty (Brodlie et al., 2012, p. 5-7).

However, user studies have indicated epistemic and political consequences when uncertainty is rendered invisible. In an analysis of approx. 100 policy makers, statistical analysts and journalists, respondents engaged with relatively simple line charts and bar charts (van der Laan et al. 2015). Confidence intervals were shown through ribbons, error bars, chisel charts, cigarette charts and boxplots. The study found that 'showing uncertainty in line charts improves the validity of the statements users make on the depicted data' (van der Laan et al. 2015, p. 231). This is specifically because when uncertainty is *not* displayed, 'respondents are overconfident in spotting a trend, even though the increase is not statistically significant' (ibid.). Users are more cautious in their assessment when a data visualisation indicates its uncertainty. Such caution is crucial in education today, given the concerns noted above about the impact of algorithmic systems on truth claims and high stakes decisions. In other user studies, data visualisations support conversations among professionals (Charleer et al., 2013). When uncertainties are shared in conversation, data visualisations become socially active and decision-making becomes collaborative (Burnett et al., 2022; Jarke & Macgilchrist, 2021), rather than 'singularising' educators as atomised data dashboard users (Decuypere & Landri, 2021).

Statisticians and members of the visualisation research community, who are also concerned with the ethicopolitical implications when data visualisations simplify, decontextualise and render certain, have argued that designs should contest these aesthetic practices (de Jonge, 2016; Tufte, 2020; Weiskopf, 2022). As we will show below, scholars and practitioners are experimenting with ways to visualise uncertainty (Levontin et al., 2020; see also D'Ignazio & Klein, 2020; in education, e.g., Alhadad, 2018; Crick et al., 2017). Overall, however, there are few experiments with visualising uncertainty in educational spaces. As a recent self-critique of the learning analytics (LA) community notes, the field is not meeting its own goals of the 'measurement, collection, analysis and reporting of data about learners and their contexts, for purposes of understanding and optimizing learning and the environments in which it occurs' (Motz et al., 2023). Despite critical research within the LA community (e.g., Ferguson, 2019) and attempts to redesign dashboards (Isaias & Backx Noronha Viana, 2020), the visualisation of these data is rarely discussed in this field (Kaliisa et al., 2023). When it is, the focus is on innovative forms of visualisation (beyond basic bar and line charts) (Charleer et al., 2014). The question thus remains: Can uncertainty be visualised in educational data dashboards, if so, how and with what (pedagogical and social) implications?

**A Critical Speculative Approach**

To explore the visualisation of (un)certainty, the following analysis draws on a critical speculative approach, inspired by social science and humanities work on speculative methods. These aim to critically interrogate the rationalities and normativities shaping current practices and thus prefiguring future possibilities, and simultaneously to develop alternative accounts, practices and sensibilities that open up ways of being/living/relating/educating otherwise (Macgilchrist et al., 2020; Ross, 2023; Selwyn et al., 2021; Wilkie et al., 2017). The goal is to unpack dominant narratives and explore possibilities for shaping other imaginaries and designing alternatives (Cerratto Pargman et al., 2022, p. 173). Speculative approaches aim to step out of 'the problem-space of the normal, the probable and the plausible', and to step into the 'speculative possibilities' that 'emerge out of the eruption of what, from the standpoint of the impasse of the present seems, in all likelihood, to be *impossible*' (Savransky et al., 2017, p. 7).

To this end, our analysis progresses in two phases, describing the need for a third phase in the concluding section. Phase 1 is a critical analysis of current LMS data visualisations. Phase 2 is an exploratory analysis of alternative designs. Phase 3 would be an active co-design process involving practitioners with complementary expertise, in this case, expertise in grassroots educational practice and expertise in the design of data visualisations.

In Phase 1, a corpus of data visualisations was created. An initial online search used terms such as 'learning management system', 'lms list', 'school lms', and 'student at risk analytics'. A snowball method expanded the online search to further platforms and visualisations for predictive analytics. This resulted in a list of 66 LMS. The websites and user documentation available online for these 66 LMS were checked manually to determine if (i) these platforms were targeted towards formal education (K-12 and/or higher education) and (ii) included predictive analytics as a feature. Platforms not fulfilling these criteria were excluded. The final corpus consisted of 19 LMS (see Fig. 1). Phase 2 consisted of desk research, exploring academic, professional and public-facing publications in which the uncertainty of predictions has been visualised. Since the first phase had found very few examples in education, this process explored further domains, identifying examples in three: defence, climate change and healthcare. In the conclusion, we propose the need for a Phase 3 to find practicable ways of visualising uncertainty through co-design with practitioners.

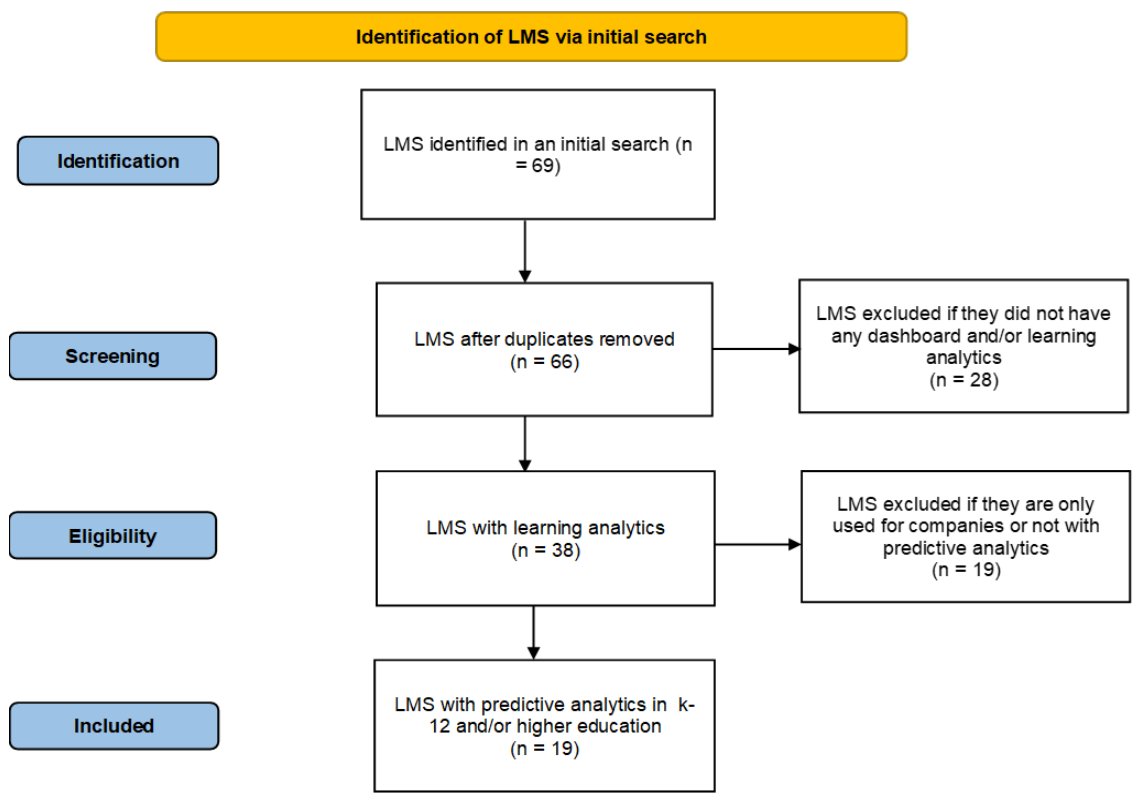

*Fig. 1. Data generation: Corpus creation*

**Phase 1. Critical Analysis: Visualising Certainty**
Analysis of the 19 sets of educational data visualisations identified three types illustrated below: traffic light colours, rainbow colours and forecasting ranges. Of the 19, only two visualised uncertainty in their risk assessment and/or predictions.[2]

*Traffic Lights*
Fig. 2, Fig. 3 and Fig. 4 illustrate a traffic light colour combination which has become classic in educational data visualisations. The risk analysis visualisation indicates low risk with green, moderate risk with yellow (amber) and high risk with red. Of the 19 LMS in our corpus, 15 used a version of the traffic light colours for assessment, behaviour and/or other indicators. In Fig. 2, the indicators are attendance, discipline and course credits, mapping to what has been called the 'ABC' of high school graduation: Attendance, Behaviour, and Course performance. Fig. 3 includes demographics alongside enrolment, engagement and academic performance as a marker of 'influence'. This dashboard also visualises the student's progress over the weeks of the course (which, in this example, falls from green through yellow to its current red status). Fig. 4 gives less insight into the basis for the visualisation, but presents risk (visualised as high, medium or low) for graduation and for postsecondary readiness, i.e., a prediction of a student's ability to be successful at college or in a career after leaving school.

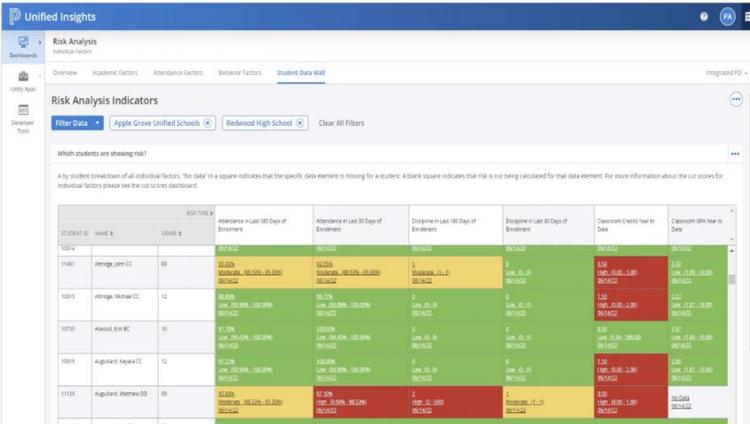

*Fig. 2. Risk analysis visualisation from Powerschool (https://www.powerschool.com/classroom/schoology-learning/)*

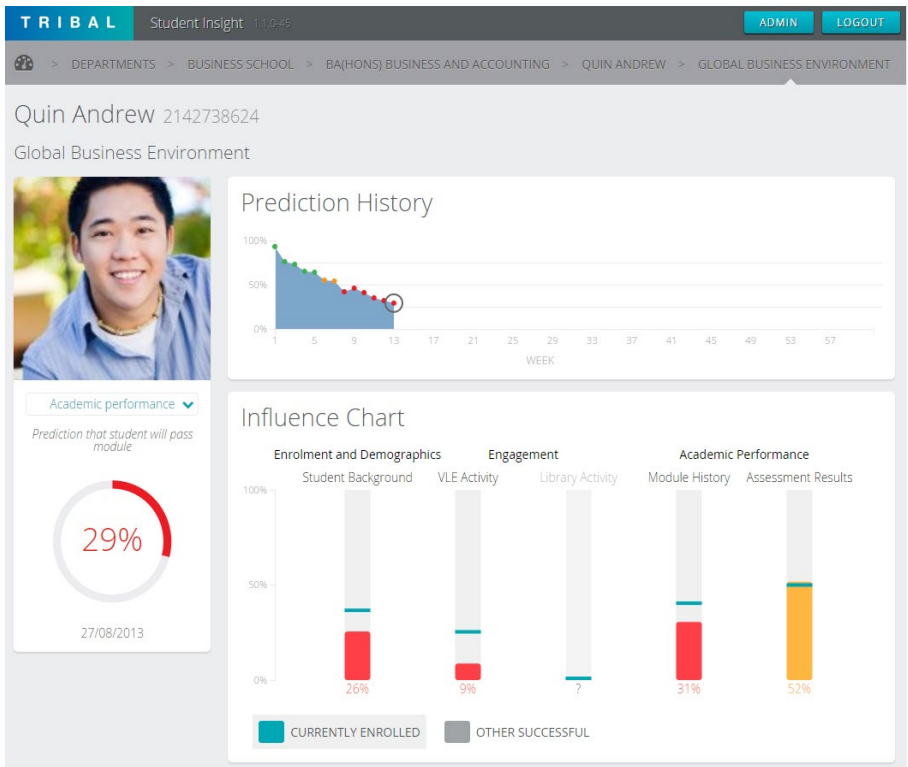

*Fig. 3. Prediction visualisations by Tribal (https://www.tribalgroup.com/)*

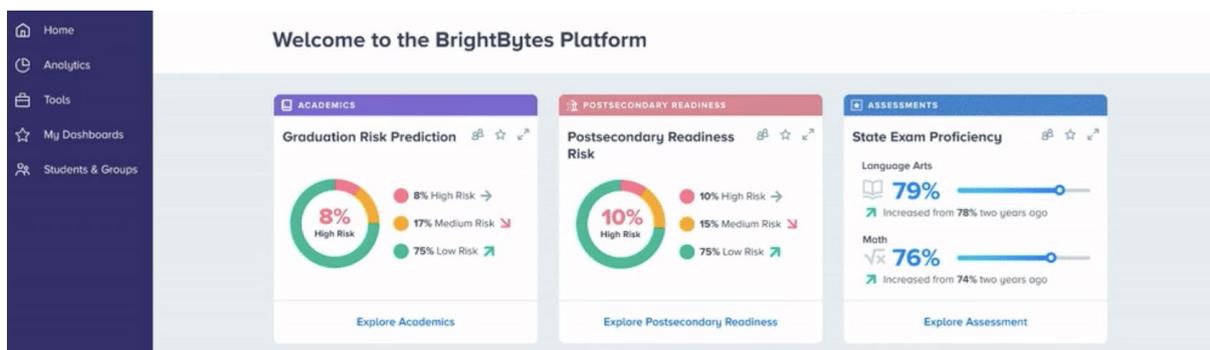

*Fig. 4. Risk prediction visualisation by BrightBytes (https://www.brightbytes.net/)*

***Rainbow Colours***

The classic traffic light colours are extended in Canvas (Fig. 5) to render a scale in rainbow colours (from red through orange, yellow and green to blue). Red continues to signal a warning, with blue signalling the highest/best grades, attendance and discipline. The linearity of the traffic light coding is also retained, with the dashboard visualising a sliding scale from risk to success. Educators can create a 'Watchlist' for students that includes up to four indicators (attendance, discipline, grades and courses). The Watchlist shows real numbers and is not colour coded.

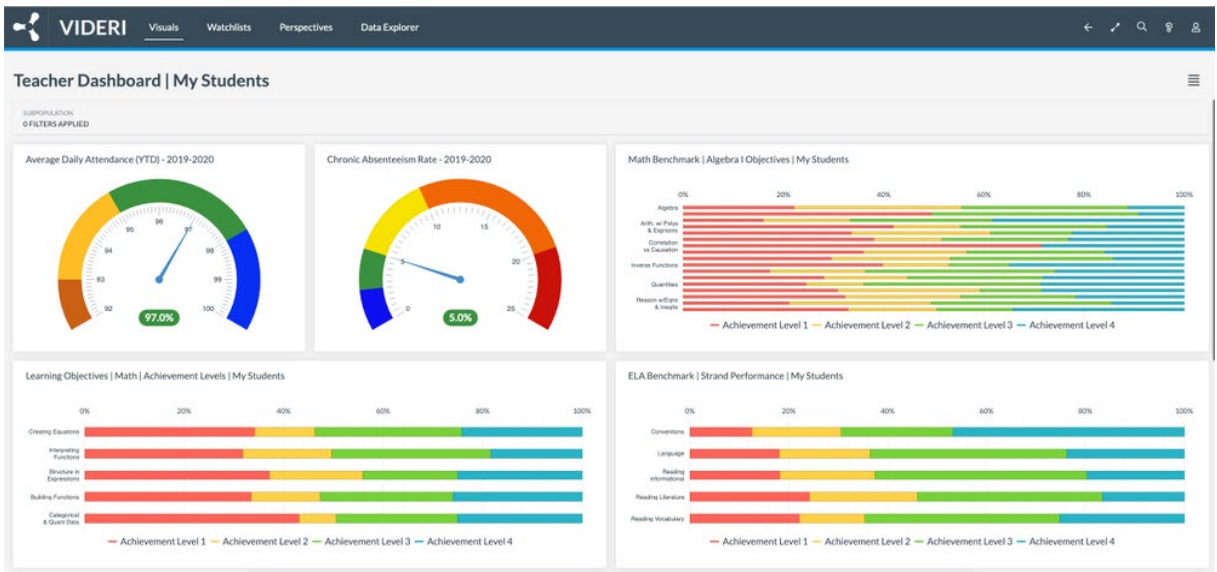

*Fig. 5. Teacher analytics dashboard in Canvas (https://www.instructure.com/canvas)*

***Projection Ranges***

A range of possible projected future outcomes is visualised in two of the systems we analysed. Blackboard (Fig. 6) visualises university students' forecasted future grades. It shows three outcomes, growing from the visualisation of their currently graded assignments (black/circles): if they continue with similar grades (orange/circles), if their grades deteriorate (to a minimum grade; red/triangles) and if they improve (to the maximum possible grade; green/rectangles).

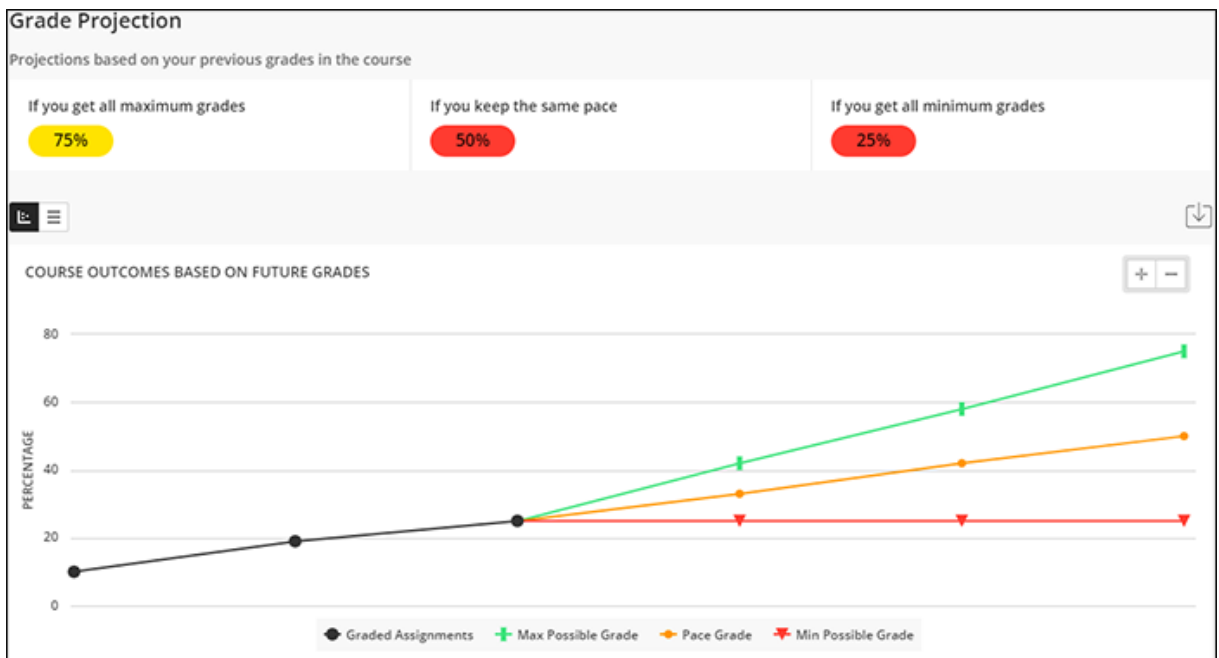

*Fig. 6. Blackboard grade performance forecasting for higher education students (https://www.blackboard.com/en-eu/teaching-learning/learning-management/blackboard-learn)*

Similarly, Fig. 7 illustrates part of a dashboard for school leaders showing school performance. The future forecasting of student enrolment numbers for the school grows from this year and the previous two years. It is shown as a brown line, marked 'forecast' with upper and lower bounds of maximum and minimum possible projected numbers showing the confidence intervals in a yellow cloud shape.

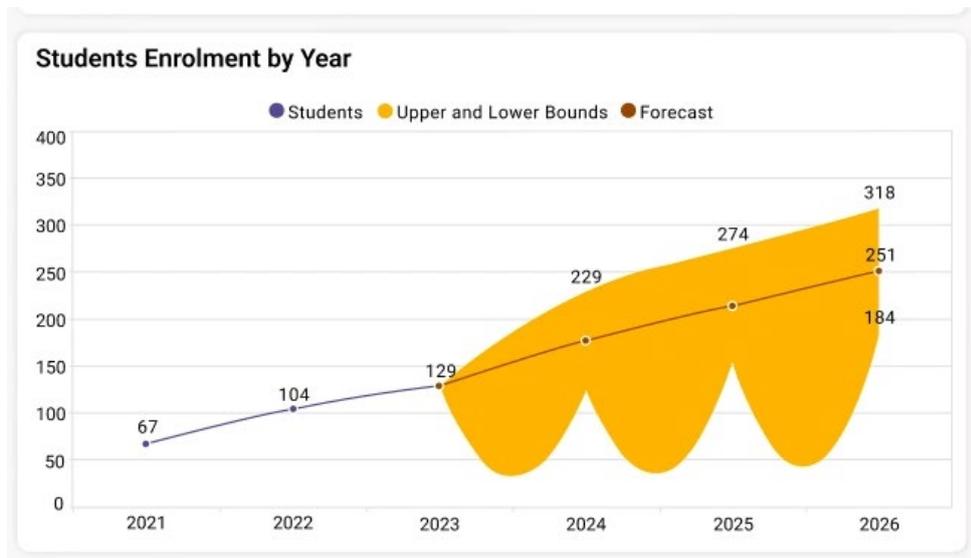

Fig. 7. Bold bi school performance visualisations for school leaders (https://www.boldbi.com/)

Interestingly, neither data visualisation operating with uncertainty and range in their forecasting are oriented to teachers. Fig. 6 is provided to students in higher education; Fig. 7 is for school leadership. We will return to the issue of developers' expectations of users below.

**Phase 2. Alternatives: Visualising Uncertainty**
Phase 2 of our speculative research design aims to enable discussion of how uncertainty is visualised in predictive systems in other social domains. Our search identified examples in defence, climate change and healthcare. We suggest that these visualisations are based on the recognition that uncertainty is an inherent part of (professional) practice and knowledge production in all three domains. This phase of analysis hopes to provide inspiration for the visualisation of uncertainty in the educational domain, if designers of predictive learning analytics (begin to) share this recognition for the educational domain.

***Defence: Simplicity and the Shared Assumption of Limited Knowledge***
In the area of defence, military personnel are often required to make critical decisions and develop strategies under time pressure and with limited information. Situations in which defence personnel operate are thus often underdetermined and inherently uncertain, for example with respect to the operating environment or the intent, capabilities or location of adversaries. Critical security scholarship has focussed on the sociotechnologies that aim to reduce insecurities and uncertainties, e.g., by anticipating 'unknown unknowns' (Daase & Kessler, 2007), through increasing tracking and targeting (Suchman et al., 2017; Weber, 2016). Interesting for educational research are the reflections on uncertainty in visualisations in a recent report by the Australian Department of Defence: 'it is important that commanders understand the associated uncertainties so that they can understand and mitigate the operational risks involved in this inherently risky enterprise' (Chung & Wark, 2016, p. 3). The report acknowledges and emphasises the *importance of staying with uncertainties* and improving decision-making by acknowledging the inherent uncertainties in which users of technical systems operate: 'Any uncertainty must be captured and communicated to aid the decision-making process' (Raglin et al., 2020, p.131). This means that uncertainties are not excluded or ironed out but purposefully foregrounded in the visualisations.

Drawing on research on users' understanding of uncertainty visualisation, the Australian defence report considers how users can be supported in understanding different levels of uncertainty. One of their examples highlights the possibility of degraded and blended icons to indicate whether radar contacts were hostile or friendly (Fig. 8).

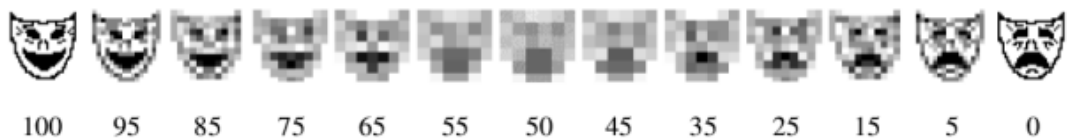

*Fig. 8. Icons representing a range of probabilities (hostile or friendly) (in Chung & Wark, 2016, p. 25, citing Finger & Bisantz, 2002)*

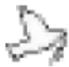

*Fig. 9. Icons showing probabilities through degrading only, or with numbers (in Chung & Wark, 2016, p. 25, citing Finger & Bisantz, 2002)*

The report also cited research findings on the effects when the degraded images were combined with numerical probability estimates (Fig. 9). They found that participants using only the icons (without numbers) showed significantly better performance on related decision-making tasks.

Interesting for educational research and practice is, firstly, the simplicity of the visualisations in these examples, and secondly the assumption in the field that professionals - with more or less tactical decision-making experience, from marines to officers - will understand the need to make decisions on the uncertain ground created by limited knowledge about the situation. In contrast to the edtech industry or ministries of education, the Australian Department of Defence draws on empirical research on how users engage with the statistical uncertainty associated with limited knowledge in order to develop simple ways of depicting this uncertainty in visualisations. They create domain-specific, high-stakes data visualisations based on the expectation that military professionals will be able and need to deal with probability and uncertainty as part of their professional practice.

***Climate Change: Multiple Models and Impact Storylines***

The lemma 'uncertain*' appears 2652 times on the 2409 pages in the 2021 report of the Intergovernmental Panel on Climate Change (ipcc, 2021). One of the reasons why uncertainty is so central to climate change and models is that 'the magnitude of feedbacks between climate change and the carbon cycle becomes larger but also more uncertain in high CO2 emission scenarios' (ipcc, 2021: 20). These uncertainties are 'dominated by the differences between emission scenarios' and in relation to the 'low-likelihood' of 'high-impact' outcomes and tipping points. Uncertainty in climate science and modelling is hence integral to the field, not only to visualise the uncertainties of climate model projections within academia but also for public science communication (Kaye et al., 2012).

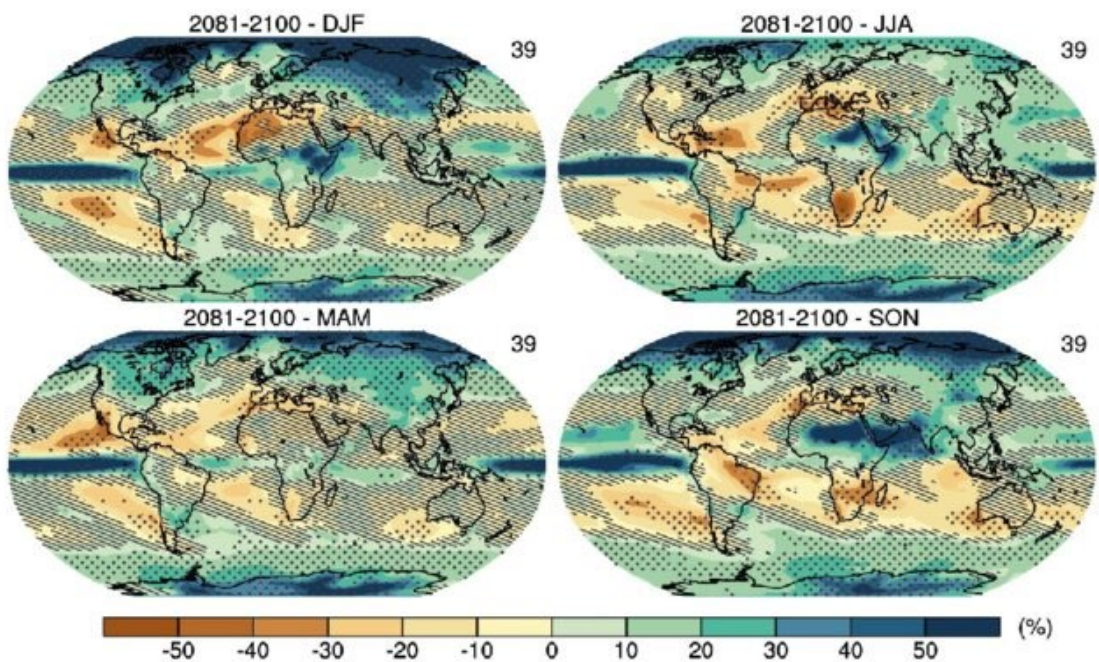

*Fig. 10. Visualising different models for seasonal mean precipitation (in Maraun, 2023)*

One way of depicting uncertainty is to visualise where regional climate models differ geographically. Fig. 10 visualises the projections of seasonal mean precipitation based on 39 models (Maraun, 2023). In each of the four images, multiple models are layered on top of one another. Hatching is used to indicate regions with little internal climate variability, i.e., where the models are in agreement. Stippling indicates regions where internal climate variability is large, i.e., where the models differ substantially.

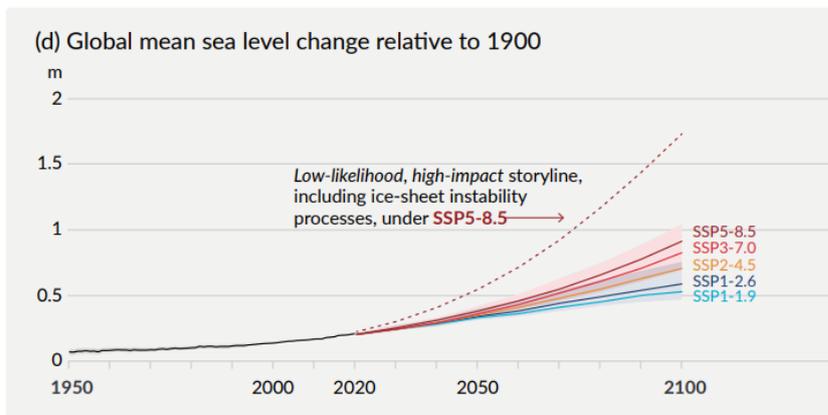

*Fig. 11. Visualising different predictions for global mean sea level change (in IPCC, 2021)*

Central to these images, with implications for education, is their visualisation of different degrees of uncertainty, each indicating which criteria play into the predictions. This shares with users the importance of accounting for uncertainty when generating data visualisations. It also shares with them the impact of collective, political action in shaping the future. Similar to the defence example above, these data visualisations address users as people who are able to assess uncertainty - based on their interest or professional expertise - and make appropriate decisions. In addition, the visualisation comments on itself as presenting narratives ('storyline'), rather than as, for instance, objective, plausible futures. Human interpretation and decision-making are implied in the data visualisation, since the collective actions of policymakers and other decision-makers will impact which prediction unfolds in the world. Transferring this to education, data visualisations would assume that educators are able to assess uncertainty in data visualisations and reflect on diverse storylines. The visualisations would indicate multiple potential outcomes that are dependent on different (collective, political, not only individual) actions (see Fig. 6).

***Healthcare: Uncertainty Language and the Response to Stress***
In the area of healthcare, professionals as well as patients experience and work under high levels of uncertainty. These relate to uncertainties about diagnostics but also more mundane aspects such as waiting times in emergency departments.

For example, Katie Walker et al. (2022) report on a co-design project in the healthcare domain, in which they developed a system for visualising waiting times in emergency departments. The authors argue that uncertainty about waiting times adds further stress to people who are already facing physical and emotional challenges. They explore how these uncertainties can be visualised in a way that is meaningful and supportive to a group of users with very diverse needs (see Fig. 12). Although operating with different proportions of text and visuals, both images in Fig. 12 explicitly include the language of uncertainty ('may be waiting', 'will most likely see').

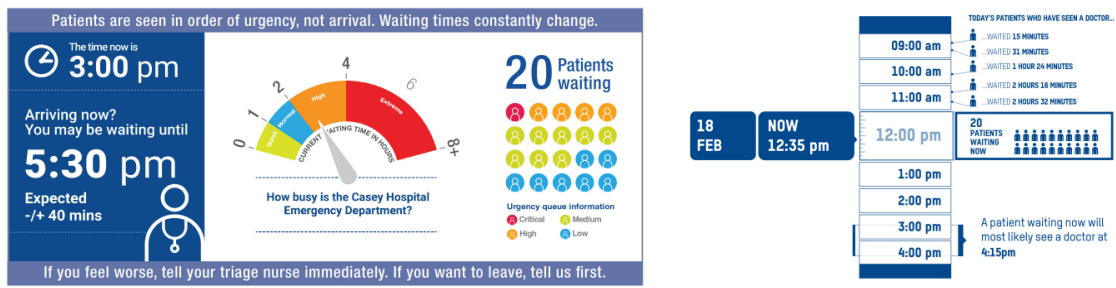

Fig. 1. Following a Rapid Iterative Testing and Evaluation (RITE) methodology, we obtained qualitative feedback from actual waiting patients on a variety of visualisation elements and styles in our study. These visualisations ranged from relatively straightforward displays relying on textual information (left) to displays replacing text with visual explanation as much as possible, for example, using a spatial representation of time and the arrival and wait time information upon which the estimated wait time window is based (right).

Fig. 12. Visualising uncertain waiting times (in Walker et al., 2022)

Key for educational research and practice is, first, Walker et al.'s (2022) acknowledgement that uncertainty can add stress, and second, their inclusion of explicit uncertainty language. Similar to the examples above, the basic assumption of visualising uncertainty here is that users can respond to and cope better with the inherent uncertainty of their situation when provided with algorithmic systems that visualise uncertainty rather than omit it. Uncertainty in education may, according to educational theory, be inherent and unavoidable, it may also be welcomed, but certain kinds of uncertainty are also, for many practitioners and students, destabilising, stressful and unwelcome. Walker et al.'s co-design research aimed to work with these potential challenges in order to find ways of visualising which show, rather than hide, the uncertainty in waiting times, and thus reduce stress. Thus, if educational data visualisations are to be designed to show uncertainty, this approach from healthcare suggests the utility of including practitioners in creating designs which support educational practices, rather than generating additional stress.

**Discussion and Concluding Thoughts**
The background to this paper is the body of work showing that data visualisations matter in education, because they impact decisions about students, and these decisions can have inequitable effects on minoritised students. We have thus focused on the infrastructural aspects of designing data visualisations, with a focus on visualising uncertainty when predictions are at stake that can open up or close down futures.

The critical analysis of current data visualisations offered in 19 learning management systems utilising predictive analytics demonstrates that, despite years of critical scholarship, the majority provides simple colour coding, charts with clear lines and edges, and singular colour-coded numbers for predictions. Few visualise the ambiguities, approximations, workarounds or statistical uncertainty inherent in predictive analytics. These visualisations enact what Edward Tufte has critiqued as 'LittleDataGraphics', data visualisations which oversimplify and thus don't actually show the data. They make invisible the 'open world' complexities from which data were generated within those algorithmic systems (Suchman, 2024). Tufte asks: 'Every day a billion people look at e-maps with data densities 20 times greater than your deceptive LittleDataGraphics. […] Why assume that readers suddenly become stupid just because they're reading your research paper?' (Tufte, 2020, p. 101) (or, in our case: 'just because they're using your data dashboard').

Interestingly, the two LMS which visualise range and/or confidence intervals in their forecasting orient not to educators but to (i) university students and (ii) school leadership. Previous research has indicated that edtech developers' expectations of teachers' technical competence are quite low (Troeger et al., 2023). This may impact developers' expectations of educators' ability to deal with complex data visualisations, and lead to the kind of simplified visualisations that our critical analysis has found. This, in turn, may be self-fulfilling: 'If they [the data science team] keep creating a reassuring image of certainty, they will continue to be asked for it' (Grant, 2017, p. 20). Indeed, a reassuring image of certainty is also offered by models of competency-based teaching and learning. Bonnet and Glazier (2023) argue that these models elide the forms of uncertainty noted in our introduction. Instead, competency-based norms assume that knowledge is relatively fixed, that learning can be planned, contained and tested, and that simple data visualisations will improve pedagogical interventions. These images of certainty diverge, however, from teachers' reported experience of education (and of edtech) as uncertain, unpredictable, contextualised and complex (Macgilchrist et al., 2023). They diverge from educational theories that foreground the inherent uncertainty and 'beautiful risk' of education, when education is understood as more than measurable cognitive learning (Allert et al., 2017; Biesta, 2013).

Operating on the assumption that education entails uncertainty (whether ontological, disciplinary, curricular, pedagogical, contextual, statistical or algorithmic), we explored data visualisations in other domains in which professionals have to deal with high levels of uncertainty: defence, climate change and healthcare. A range of visualisations have been created in these domains that embrace the observation that '[t]he world is not a solid continent of facts sprinkled by a few lakes of uncertainties, but a vast ocean of uncertainties speckled by a few islands of calibrated and stabilized forms' (Latour, 2005, p. 245). For example, the visualisations from climate science demonstrate how different models represent different climate projections (hatched areas on map) or how different storylines may be depicted in a graph. Similarly, the healthcare example demonstrated how different data categories (e.g., number of patients, predicted urgency) are included in projections of waiting times and allied with a vocabulary that highlights uncertainty. In all three domains the design included the expectation that users need to, and will be able to, deal with the inherent uncertainty of the domain, of the algorithmic arrangements, and of what is likely to happen next. In this way, users are encouraged to interpret the uncertainty of predictions and re-contextualise the predictions through their own 'situated knowledges' (Haraway, 1988). In the case of education this would mean that dashboards actively encourage teachers to consider the uncertainty of predictions in relation to their situated knowledge about their students' attendance, behaviour, course performance, demographic indicators, platform interactions, as well as family situations, friendships, struggles and structural inequalities. Teachers already regularly do this, but currently they have to contest the ways data visualisations act on the world if they want to foreground their situated understandings (Hangartner et al., 2024).

We draw two sets of conclusions for educational research and practice. First, we note differences in making uncertainty visible, with consequences for education, including professional practice. The differences are related on the one hand to whether a simple graphic is created with clear-cut lines that suggest a certainty to the system's predictions of student risk and success, or more complex graphics are designed that foreground the inherent uncertainty of predictions. The differences are related on the other hand to presuppositions about professionals using the data visualisations, i.e. the extent to which the expectation is inscribed that professionals *regularly* deal with, *can* deal with and *should be supported* in dealing with various dimensions of uncertainty. The examples demonstrate how, for instance, military officers are expected to recognise the uncertainty in their domain, and understand that decision-making operates on uncertain terrain, whereas educators using LMS are expected to draw conclusions and pedagogical insights from LittleDataGraphics which have erased the uncertainties of open worlds, the uncertainties of algorithmic prediction, and the uncertainties of futures yet to be formed. Each possible form of data visualisation invites educators to relate to themselves and their students in particular ways, and backgrounds others ways of relating. This in turn shapes how decisions are made and whether (equitable) futures are opened up or closed down. These professional practices belong to the apparently banal, but by no means benign, ways in which educational values and priorities are changing.

Second, more practically, we propose implications of this analysis for design. While there are strong arguments for avoiding datafication per se and derailing algorithmic regimes entirely, in this paper we propose instead to collectively transform the development and design of datafying infrastructures (see also, e.g., Swist & Gulson, 2023; Viljoen, 2021). How can educators' lived experience of education as inherently uncertain and risky become represented in LMS' data visualisations? As the *Analysis under Uncertainty for Decision-Making Network* has observed: 'There is no 'optimal' format or framework for visualising uncertainty. Instead, the implementation of visualisation techniques must be studied on a case-by-case basis, and supported by empirical testing' (Levontin et al., 2020, p. 4). The analysis in this paper suggests several concrete visualisation techniques with which educational products could experiment. A Phase 3 of this current research would bring together data visualisation experts interested in complex visualisations and educators with expertise in classroom practice. In co-design processes, they could create practicable yet sufficiently complex data visualisations. These should aim to support decision-making which is appropriately cautious (and thus more valid) rather than over-confident (van der Laan et al., 2015). The visualisations could enable educators to reflect critically with students about prediction as one of the core features of datafied life. They could open up rather than shut down educators' degrees of freedom. This could make the inherent beautiful uncertainty of education visible – and yet not an additional stressor – in educators' professional worlds.

**Notes**

[1] Details on the indicators, also called 'predictors' in the Moodle documentation, are available here https://moodle.org/ and here: https://docs.moodle.org/402/en/Learning_analytics_indicators.

[2] A third, Octobusi, also visualised statistical range as boxplots in one of its presentations. We have not included these in our findings, however, since Octobusi does not link this visualisation to predictions, nor meaningfully connect it to other data.

**Disclosure statement**

No potential conflict of interest was reported by the authors.

**ORCID**

Felicitas Macgilchrist http://orcid.org/0000-0002-2828-4127
Juliane Jarke http://orcid.org/0000-0001-8349-2298